\documentclass[10pt,english]{article}
\pdfoutput=1
\usepackage[T1]{fontenc}
\usepackage[latin9]{inputenc}
\usepackage{geometry}
\usepackage{cite}
\usepackage[usenames,dvipsnames]{color}
\usepackage{graphicx}
\usepackage{amsmath}
\usepackage{amsfonts}
\usepackage{amssymb}
\usepackage{dsfont}
\usepackage{dcolumn}
\usepackage{bm}
\usepackage{slashed}
\usepackage{pstricks}
\usepackage{slashed}
\usepackage{multirow}
\usepackage{array}
\usepackage{rotating}
\usepackage{xcolor}
\usepackage{epstopdf}
\usepackage{fancybox}
\usepackage{ulem,fancyvrb}
\usepackage{color}
\usepackage{mathdots}
\usepackage{mathrsfs}
\usepackage{multirow}
\usepackage{esint}
\allowdisplaybreaks

\newcolumntype{x}[1]{
{\centering}p{#1}}%

\newcommand{\GeV}      {~\mathrm{GeV}}

\newcommand{\beqn}{\begin{eqnarray}}
\newcommand{\eeqn}{\end{eqnarray}}
\newcommand{\be}{\begin{equation}}
\newcommand{\ee}{\end{equation}}

\newcommand{\mathsym}[1]{{}}

\def \n34{\tilde{\chi}^{0}_{3,4}}

\def\met100{\slashed{E}_T\geq 100 \GeV}
\newcommand{\st}{Stueckelberg~}

\newcommand{\gappeq}{\mathrel{\rlap {\raise.5ex\hbox{$>$}}
{\lower.5ex\hbox{$\sim$}}}}
\newcommand{\lappeq}{\mathrel{\rlap{\raise.5ex\hbox{$<$}}
{\lower.5ex\hbox{$\sim$}}}}

\def\met{\slashed{E}_{T}}

\begin{document}

\title{\textbf{GUT and Supersymmetry at the LHC and in Dark Matter}}
\author{
Pran~Nath\footnote{Email: nath@neu.edu}
\\\textit{Department of Physics, Northeastern University, Boston, MA 02115, USA}}
\date{}
\maketitle

%%%%%%%%%%%%%%%%%%%%%%%%
%%%%%     abstract
%%%%%%%%%%%%%%%%%%%%%%%%

%%%%%%%%%%%%%%%%%%%%%%%%
%%%%%     introduction
%%%%%%%%%%%%%%%%%%%%%%%%

%\newpage

\begin{abstract}
Conventional $SO(10)$ models involve more than one scale for a complete breaking of the
GUT symmetry requiring  further assumptions on the VEVs of the  Higgs fields
that enter in the breaking 
 to achieve 
viable models. Recent works where the breaking can be accomplished at one scale
are discussed. These include models with just a 
pair of  $144+\overline{144}$ of Higgs fields. Further extensions of this idea utilizing 
 $560+ \overline{560}$ of Higgs representations allow both the breaking at one scale, as well
 as accomplish a natural doublet-triplet splitting via the missing partner mechanism.
More generally, we discuss the connection of high scale  models to low energy physics 
in the context of supergravity grand unification. 
Here we discuss a natural solution to the
little hierarchy problem and also discuss the implications of the LHC data for supersymmetry.
It is shown that the LHC  data implies that most of the parameter space of supergravity models
 consistent with the data 
lie  on the Hyperbolic Branch of radiative breaking of the electroweak symmetry and more specifically
on the Focal Surface of the Hyperbolic Branch.  A  discussion is also given of the
implications of recent LHC data on the Higgs boson mass for the discovery of  supersymmetry 
and  for the search for dark matter.  \\ \\

\end{abstract}

%%%%%%%%%%%%%%%%%%%%%%%%%%%%%%%%%%%%%%%%%%%%
%% MAINMATTER
%%%%%%%%%%%%%%%%%%%%%%%%%%%%%%%%%%%%%%%%%%%%

\section{GUTs and Supersymmetry }
{\it Intoduction:}  Grand unification ~\cite{Pati:1974yy,Georgi:1974sy,georgi} for the description of particle interactions 
is desirable for a variety of reasons~(For a review see~\cite{Nath:2006ut}).
We discuss 
 some of the recent developments  in unified models which are relevant in view of the hunt for 
new physics at the large hadron collider~\cite{Nath:2010zj}. Of specific interest is the 
gauge symmetry based on $SO(10)$~\cite{georgi} which provides a framework for unifying the $SU(3)_C\times SU(2)_L\times U(1)_Y$ gauge groups and also for unifying quarks and leptons in one generation  in a single $16$--plet spinor representation.
Additionally, the 16--plet also contains a right--handed
singlet state, which is needed to give mass to the neutrino via the seesaw mechanism. 
Supersymmetric
$SO(10)$ models have the added attraction that they predict correctly the unification of gauge couplings, and
solve the hierarchy problem by virtue of SUSY. 
 However, SUSY $SO(10)$ models, as usually constructed, have
two drawbacks, both related to the symmetry breaking sector.
 First, there are two different mass scales  involved in the breaking of the GUT symmetry, one to
reduce the rank and the other to reduce the symmetry all the way to $SU(3)_C\times SU(2)_L\times U(1)_Y$. 
 Thus typically three types of Higgs fields are needed:
one  for rank reduction such as $16+\overline{16}$ of $126+ \overline{126}$ and then
a $45$, $54$ or  $210$ for breaking the symmetry down to the standard model symmetry, 
 and a $10$ plet  for electroweak symmetry breaking.
Second, the  GUT models typically have the so called  doublet-triplet problem where
one needs an extreme fine tuning to make the Higgs doublets light. 
These drawbacks can be corrected in a new class of models recently proposed~\cite{Babu:2005gx,Nath:2005bx}. 
Here we will review first this class of models.  We will then go on to discuss a variety of other topics
which connect high scale models to low energy physics. These include a discussion of supergravity 
grand unification, the little hierarchy problem,  the implications of the  recent data from ATLAS and
CMS on the Higgs boson. Other topics discussed include proton decay in GUTs,
and dark matter. Finally we will discuss a topic which is not necessarily tied to a GUT theory but is of 
importance and of current interest, which is cosmic coincidence, i.e., the fact that 
dark matter and baryonic matter  are roughly in the five to one ratio. We will discuss a possible solution to 
this within the framework of a \st extension of the standard model and of the minimal supersymmetric
standard model (MSSM). We discuss below the 
topic listed above in further detail.
\\

{\it One step breaking of GUT symmetry:}
Multiple  step breaking requires  additional assumptions  relating VEVs of  
different Higgs representations 
to explain the  gauge coupling unification of the electroweak and the strong interactions
(for a review see ~\cite{Dienes:1996du}) reducing predictivity, while such an assumption is unnecessary
in a single step breaking~\cite{Babu:2005gx,Nath:2005bx}. 
In $SO(10)$ the simplest combination of  Higgs representations that can achieve a single step breaking is
 $144+\overline{144}$. 
  The reason for this is easily understood by looking at the decomposition of the $144$ plet of $SO(10)$  under $SU(5)\times U(1)$ so that 
 $\overline{144}  
= \bar 5(3) + 5(7) + 10 (-1) + 15 (7) + 24(-5) + 40(-1) + \overline{45}(3)$.
Here the $24$ plet carries a $U(1)$ quantum number and thus a VEV formation of it will
reduce the rank of the group as well as break $SU(5)$.
Additionally one can obtain a pair of light Higgs doublets needed for electroweak symmetry breaking 
 from the  same irreducible $144+\overline{144}$ Higgs multiplet.
Thus one can achieve the full breaking of 
 $SO(10)$ into $SU(3)_C\times U(1)_{em}$  by just one pair of $144+\overline{144}$.
Dealing with the $144$ representation requires special techniques ~\cite{ns} which utilize oscillator
method~\cite{ms} (For alternate techniques for the computation of $SO(10)$ couplings see 
~\cite{Fukuyama:2004ps,Aulakh:2004hm}).
With a  $144+\overline{144}$ of Higgs fields the
    first two generations of fermion masses arise from quartic couplings 
$(1/\Lambda) 16.16.144.144$ and $(1/\Lambda)16.16.\overline{144}.\overline{144}$
where $\Lambda\sim O(M_{Pl})$ and since the effective Yukawas  are scaled by  $<144>/\Lambda$
and $<\overline{144}>/\Lambda$ where $<144>=<\overline{144}>\sim O(M_{G})$ 
the Yukawas for the first two generation fermions are naturally small.
 However, additional matter representations are needed for generating large third generation masses.
 Such additional representations may be $10$, $45$ or $120$ of matter fields.
 In models of the above type the third generation masses involve very significant representation mixing.
 It is then possible to  modify  the conditions for $b-t-\tau$ unification to occur. Specifically
 one finds that such a unification can come about for low values of $\tan\beta$, i.e., a $\tan\beta$ as
 low as 10 in contrast to  models where the Higgs doublets arise from the $10$ plets and  
 a $b-t-\tau$ unification requires a  large $\tan\beta$~\cite{Ananthanarayan:1991xp}.\\
    
   {\it The doublet-triplet problem:} The second problem in  GUT theories typically concerns  the doublet-triplet mass splitting, i.e.,  one must do an extreme fine--tuning at the level of one part in $10^{14}$ to get the Higgs doublets
of MSSM light, while  the color triplets-anti-triplets  remain superheavy.  Some possible solutions to the doublet -triplet problem include:
 (i)  The missing  VEV solution where  SO(10) breaks in the B-L direction which allows the Higgs triplets 
 and anti-triplets 
 to be
 heavy while the Higgs doublets remain light~\cite{DW}; (ii)  The  flipped $SU(5)\times U(1)$; (iii) The 
   missing partner mechanism,   and (iv)
 Orbifold GUTs~\cite{Kawamura:2000ev}. The missing partner mechanism  and  the orbifold GUTs are rather  compelling in that some 
doublets are forced to be massless. We will focus on the missing partner mechanism~\cite{SU(5)Missingpartner}
and  discuss how it works in 
$SU(5)$ and then discuss how one can extend to $SO(10)$.
   In $SU(5)$ the Higgs sector consists of $50, \overline{50}$ and $75$ plets of heavy  representations 
   and  $5, \bar 5$ of light representations. The $75$ plet breaks the GUT symmetry, while
   $50, \overline{50}$ have the property that they contain a Higgs tripet- anti-triplet pair but no
   higgs doublet pairs. When mixing is introduced between the light and the heavy sectors, 
   the Higgs triplets-anti-triplets become heavy while the doublets remain light. Thus here one naturally achieves a
   pair of light higgs doublets.  \\

   In $SO(10)$ the missing partner mechanism is more complex. This complexity arises due to 
   the additional problem of exotics. The reason for this new feature in $SO(10)$ is due to the light
   sector which unlike the light sector in the $SU(5)$ case contains  a new array of massless fields
   which must also be made heavy. One example of a missing partner mechanism in $SO(10)$ was achieved
  in ~\cite{SO(10)Missingpartner}
    consisting of a heavy sector with 
 $126+\overline{126} + (210+16+\overline{16})$ of Higgs and  a light sector with
  $10+120$ plet of Higgs. The  $210+ 16+\overline{16}$ fields are responsible for the breaking of the
  $SO(10)$ gauge symmetry. The remaining heavy sector fields, i.e.,  
  $126+\overline{126}$ fields do not mix with the $210+ 16+\overline{16}$ and do not acquire VEVs.,
  but they play the same role that $50+\overline{50}$ play in $SU(5)$, i.e., they contain 
   an excess of triplet and anti-triplet pairs relative to the doublet pairs. After mixing with the light 
   fields a careful count shows that there is one pair of Higgs doublets which are left massless while
   all the Higgs triplets and anti-triplets gain mass. Thus the missing partner mechanism works. More recenlty
  several more cases have been discovered~\cite{Babu:2011tw}.
  These include one model where the heavy sector is
 $126+\overline{126}+ 210$ and the light sector consists of   $2\times 10+120$, while another model
 consists of  a  heavy sector with $126+\overline{126}+ 45$ and a light sector consisting of
  $10+120$.    In addition an entirely different array of representations have been found which
  also lead to the missing partner mechanism. This  possibility consists of
 $560+\overline{560}$ plets which constitute a heavy sector and  $2\times 10+ 320$ plets which constitute 
 a light sector. 
 Here $560+\overline{560}$ contains an excess of Higgs triplet and anti-triplet pairs over Higgs doublet pairs
 which again allows for the missing partner mechanism to work. 
This is easily seen by decomposing the $560$ and the $320$  multiplets into their $SU(5)$ components. Thus  
$ 560= 1(-5) + \bar 5 (3) + \overline{10}(-9) + 10(-1) + 10(-1) + 24(-5)
+ 40(-1)  +45(7) + \overline{45}(3) + \overline{50}(3) + \overline{70}(3)
+ 75(-5) + 175(-1)$ and  for the $320$ multiplet one has 
$ 320 = 5(2) + \bar 5(-2) + 40(-6) + \overline{40} (6) + 45(2) + \overline{45}(-2) + 70(2) + \overline{70}(-2)$.
A simple count of the Higgs doublets and triplet-anti-triplet pairs shows that the heavy sector consisting of 
$560+\overline{560}$ has 4 doublet pairs and five triplet-anti-triplet pairs, while the light sector consisting 
of $2\times 10+320$ multiplets consist of  5 doublet pairs and five triplet-anti-triplets pairs leaving us 
one light doublet pair and no light Higgs triplet -anti-triplet pair after mixing of the light and the heavy sectors. 
The multiplets $560$ and $320$ have not surfaced in GUT model building before but arise
as  natural possibilities in the context of achieving a missing partner mechanism in $SO(10)$.
The models of the type discussed above have the possibility of accommodating  large neutrino mixing angles
and also relatively large  values of $\theta_{13}$ as indicated by the recent data from experiment~\cite{An:2012eh}. 
The detailed implications of these models still need to be worked out. \\

One possible danger in the missing partner mechanism concerns the  
 possibility  that the Higgs doublets may receive contributions of size $M_G^2/M_{Planck}$ 
arising from  Planck scale corrections. The presence of such corrections would spoil  
 the missing partner mechanism.  However, as discussed., e.g., in~\cite{SO(10)Missingpartner,Dvali:1996sr},
  such corrections can be forbidden by an  anomalous $U(1)$ symmetry. 
  Another issue concerns the nature of the theory above the scale $M_G$. Thus  
 because of the large number of degrees of freedom,  models of the above type are not asymptotically
 free above the unification scale $M_G$.  However, the region above $M_G$ is the region where
 gravity becomes strong. In particular this happens when there are  a large number of degrees of freedom
 $N$ involved. As pointed out recently ~\cite{bdv} in this case the effective fundamental scale 
 is reduced by a factor $\sqrt{N}$ which lies in the vicinity of $M_G$. Because of the proximity of the
 fundamental scale to  $M_G$, the effect of 
 non-renormalizable interactions would be very significant and
 must be included above the scale $M_G$.  Inclusion of such terms would redefine the  theory 
 in a significant way above $M_G$ and the appropriate procedure in this region then is to use here
 the UV complete theory rather than a truncated version of it.\\

 {\it SUGRA unification and LHC implications:}  In order to make contact with low energy physics one needs to break supersymmetry in a supersymmetric grand unification. 
 It is difficult to achieve such a breaking within global supersymmetry and one must make supersymmetry 
 a gauged symmetry which brings in gravity~\cite{anz} and thus gravity enters 
  in an intrinsic 
 way into model building. Specifically we need the framework of applied $N=1$ 
 supergravity ~\cite{appliedsugra,Cremmer:1982en} to build 
 models and in particular  supergravity grand unification\cite{can}. 
 A broad class of models fall under this
rubric. These include mSUGRA (sometimes referred to as CMSSM)~\cite{can}, and SUGRA models with non-universalities in the
Higgs sector and in the gaugino sector
(see
~\cite{Nath:1997qm,Chattopadhyay:2001mj} and the references therein).
Non-universalities of soft masses also arise in string models and D brane models (see~\cite{Kors:2003wf}
and the references therein). 
 The parameter space of mSUGRA is well known consisting of 
 $ m_0, m_{1/2}, A_0, \tan\beta, {\rm sign}(\mu)$
 where $m_0$ is the universal scalar mass, $m_{1/2}$, the universal gaugino mass, $A_0$ the universal
 trilinear parameter all taken at the GUT scale, while $\tan\beta=<H_2>/<H_1>$ where $<H_2>$ gives mass to 
 the up quarks, and $<H_1>$ gives mass to the down quarks and  the leptons, and $\mu$ is the Higgs mixing
 parameter in the superpotential. 
For non-universal SUGRA models there are additional parameters. For example, for the gaugino sector
one can choose different gaugino masses at the grand unification scale, i.e., 
$ \tilde m_1, \tilde m_2, \tilde m_3$  for the $U(1), SU(2), SU(3)$ sectors, and 
$m_{H_1}, m_{H_2}$ for the Higgs sector.  
There are also other mechanisms for the breaking of supersymmetry such as 
 gauge mediation, anomaly mediation and the breaking by an anomalous $U(1)$ 
 as well as 
 involving mixtures of these such as a  mix  of gravity breaking and anomalous 
 $U(1)$ breaking~\cite{Kors:2004hz}. 
 Comparison with low energy data, of course,  requires renormalization group evolution 
 (see, e.g., ~\cite{Martin:1993zk} and for a review see ~\cite{Ibanez:2007pf} and
  the references therein)
   to compute the sparticle mass spectrum.
 Now the sparticle landscape is rather large~\cite{landscape} and one needs tests at colliders
 to delineate the nature of soft breaking using experimental data (for a review see ~\cite{Nath:2010zj})
 and for related works see ~\cite{msugrarelated,relatedanalyses_landscape}).   
 The sparticle spectra can be affected by CP violating effects (for a review see ~\cite{Ibrahim:2007fb}).
 One of the important signatures for supersymmetry is the trileptonic 
 signal ~\cite{Chamseddine:1983eg,Dicus:1983cb}. This signature from the 
  on-shell decay of the $W^{\pm}\to \tilde \chi^{\pm}\chi_i$ at colliders was discussed in ~\cite{Baer:1986vf},
 and for the off-shell decays in  ~\cite{Nath:1987sw} with further work in  ~\cite{Baer:1992dc} 
 and in several other papers.\\
 
{\it The little hierarchy problem:}
One version of the so called  little hierarchy problem relates to keeping  $\mu$ small while $m_0$ gets  large.
Using radiative  breaking of the electroweak symmetry one can write
~\cite{Chan:1997bi,HB2,bbbkt,Feldman:2008jy,Akula:2011jx}
(for related works on naturalness see~\cite{FMM,naturalness,Feldman:2011ud})
~$ \mu^2  = -\frac{1}{2}M_Z^2 +  m^2_0  C_1+ A^2_0 C_2 +
 m^2_{\frac{1}{2}} C_3+ m_{\frac{1}{2}}
A_0 C_4+ \Delta \mu^2_{loop}$.
The  case  $C_1>0$  is the so called Ellipsoidal Branch.  Now  in certain regions of the parameter space $C_1$
can vanish or  turn negative. This converts the REWSB equation from  an Ellipsoidal Branch to a Hyperbolic
Branch (HB).
 One can further classify the HB region into three  separate regions~\cite{Akula:2011jx}. These are: (i) The Focal Point region
 HB/FP where $C_1=0$; (ii) The Focal Curve  region  HB/FC: Here 
 $C_1<0$ and  two soft parameters can get large for fixed $\mu$, and  (iii) The Focal Surface region 
HB/FS:  Here $C_1<0$ and three soft parameters $m_0, m_{1/2}, A_0$ can get large for
fixed $\mu$. 
Now from the RG analysis it is possible to write $C_1$ in the form $C_1=(1-3D_0)/2$ where 
$D_0(t) = \exp[-6\int_0^{t} Y_t(t') dt']$ and $Y_t= h_t^2/4\pi^2$ where $h_t$ is the top Yuakwa coupling
and where $t=log(M_G^2/Q^2)$. It is also easily seen from the
solution to the RG equations that the RG correction to the Higgs boson mass that couples to the 
top quark, i.e., $H_2$, is given by $\delta m^2_{H_2}  = m_0^2 (3D_0-1)/2$ and is thus related to
$C_1$ simply as $\delta m^2_{H_2} \simeq-m_0^2 C_1$.
 One finds then that $C_1$ vanishes for the case when $D_0=1/3$ which also implies the vanishing of the
 correction $\delta m^2_{H_2}$. This is the Focal Point region (HB/FP). Recent analyses of the
LHC data within supergravity grand unification with universal boundary conditions
show that the HB/FP region is mostly depleted and the bulk of the region which remains 
lies in the focal curve HB/FC and focal surface region HB/FS~\cite{Akula:2011jx}. \\

 {\it The Higgs boson mass:} 
 In SUGRA models at the tree level the mass of the light neutral CP even  
  Higgs  boson mass is less than $M_Z$.  However, it 
 can be lifted above $M_Z$ by loop corrections (For a review see~\cite{Carena:2002es}).
 Specifically in  SUGRA   models  the light Higgs boson mass can run up to around 130 GeV 
  with $m_0$ in the TeV region~\cite{Akula:2011aa}. Recent data gives some positive hints for a Higgs boson
  mass in the vicinity of around 125 GeV~\cite{dec13} which offers support for the supergravity unified  model.
  We note that a Higgs boson mass in the region around 125 GeV points to a heavy sparticle spectrum
  specifically heavy scalars some  of which could be several TeV in mass (see also in this context  ~\cite{Wells:2004di}    
  where a PeV scale has been discussed). However, many sparticles could still be light, i.e.,  the light stop and the
  sbottom, charginos and neutralinos. Further, a  gluino mass in the vicinity of  a TeV is still
  allowed ~\cite{Akula:2011aa}.  
  Detailed analyses of the sparticle spectrum and of the Higgs sector require imposition of the experimental constraints.
  These include the relic density constraint on the neutralino dark matter, the constraints from the flavor changing processes
  $b\to s+\gamma$~\cite{Degrassi:2000qf}, $B_s\to \mu^+\mu^-$~\cite{bmumu},
  from the muon anomalous magnetic moment~\cite{Yuan:1984ww}, 
  as well as  constraints from
  the sparticle lower limits  from experiment ~\cite{pdgrev}. We note in passing that
  the Higgs boson sector is affected by CP phases and new detectable  phenomena arise due to mixing of the
  CP even and CP Higgs bosons~\cite{pilaftsis}.  
  \\

{\it Proton decay:}
Another aspect of grand unification is 
proton decay which is a generic feature of unified models of particle interactions.
There are a variety of sources for proton decay.
The most generic feature in GUT models is 
proton decay via the exchange of vector lepto-quarks. 
The current experimental limit $\tau (p \to \pi^0 e^+) \ > 1.4 \times 10^{34}$ yrs
implies a  very rough lower bound on the superheavy gauge boson mass of
$M_V \ > \ 5 \times 10^{15} \ \ \textrm{GeV}$. 
Thus  proton
stability at current levels  implies  the existence  of a very high scale,  much closer to the Planck scale
than the weak scale.  
In supersymmetric grand unification further constraints are needed.
Thus 
$B\&L$  violating dim 4 operators  appear in SUSY, i.e., $Q L D^C, U^C D^C D^C, LL E^C, LH,$
leading to fast proton decay and their suppression requires R parity. 
However, even with R parity B\&L violating dimension five operators, i.e., 
$QQQL/M_T$, and
$ U^CU^CD^CE^C/M_T$, are allowed and generate proton decay. Thus 
dressing loops convert  these dim 5 operators to  B\&L violating dim 6 operators involving  quarks and leptons. Further, these
dim 6 operators are converted to an effective lagrangian  involving mesons and baryons 
 which allow one to calculate proton decay processes such as 
$p\to \bar \nu_{e,\mu,\tau} K^+, ~\nu_{e,\mu,\tau} \pi, ~\nu_{e,\mu,\tau} \eta,  ~\mu \pi, ~eK, ~\mu K$.
A variety of GUT and string models can be tested using proton decay constraints~\cite{Arnowitt:1993pd}.
There is significant  model dependence in the predictions of the proton decay modes
  specifically of the SUSY decay modes.
   Thus the decay lifetimes depend on the 
   nature of soft breaking which enters in the  dressing loop diagrams~\cite{enr,pdecay,Lucas:1996bc}.
Additionally proton decay can be affected by: (i) gauge coupling unification which constrains the 
 Higgsino triplet mass; (ii) quark-lepton textures, (iii) FCNC and dark matter constraint,  and (iv)  gravitational warping effects~\cite{Hill:1983xh,Dasgupta:1995js}. Finally we may mention that the 
accuracy of effective  lagrangian  approximation which converts 
operators such as $QQQL$  and $U^CU^CD^CE^C$ into lagrangian with mesons and baryons
can have an effect on the proton decay life time predictions.  
Further, predictions of proton lifetime for supersymmetric decays  depend significantly on  
 the particulars of the supersymmetric  grand unified model. 
Often a  suppression of B\&L violating 
dimension five operators is needed to raise the proton lifetime beyond the current experimental
limits. 
A variety of possibilities exist  for such a  suppression. 
These are: (i) The cancellation mechanism ~\cite{Nath:2005bx} which can 
 operate if there are  more than one source of Higgsino mediation, (ii) 
  One or more couplings that enter in Higgsino mediated proton decay 
 are naturally suppressed  by an internal mechanism, and (iii) 
$m_0$ is large and REWSB is realized on the Hyperbolic Branch where large 
scalar masses consistent with small $m_{1/2}$ and $\mu$ are allowed.
An example of the  cancellations or  suppression of Higgsino mediated proton decay 
occurs in the $144+\overline{144}$ Higgs model~\cite{Nath:2005bx}. 
Proton decay lifetime  has been calculated in the $144+\overline{144}$  unified higgs model
and the results are consistent with current experiment and allow for the possibility of the discovery of
the decay  in improved p decay experiments~\cite{Wu:2009zzh}.   One important conclusion that one finds in this and
other similar analyses is that the discovery of proton decay may be just around the corner and such
a discovery may occur even with modest improvements in the sensitivity of proton decay experiment. \\

{\it Gauged B-L, R parity  and the Stueckelberg mechanism:}
As can be seen from the preceding discussion,
R-parity (defined by $R=(-1)^{2S + 3 (B-L)}$, where $S$, $B$ and $L$ stand for the spin, baryon and lepton numbers, respectively) is an important symmetry in supersymmetric theories and is often 
imposed on GUT models on phenomenological grounds but it  could also 
be automatic  ~\cite{Martin:1992mq}.
However,  R parity even if 
preserved at the GUT scale can undergo spontaneous breaking due to renormalization group effects
(see, e.g., ~\cite{Barger:2008wn}  and the references therein). It is then interesting to ask 
if the radiative breaking of  R parity can be evaded. One possibility is that R parity arises as a remnant 
of a gauged $U(1)_{B-L}$. Thus within MSSM one may have an anomaly free $U(1)_{B-L}$ by
 inclusion of right handed neutrinos,  one for each generation. 
 Also GUT models such as $SO(10)$ and $E_6$ and string models possess  a 
 gauged $U(1)_{B-L}$. However, the massless gauge boson associated with a gauged $U(1)_{B-L}$ 
 must grow a mass to evade generating an undesirable long range force. The simplest way to 
 accomplish this is by the Stueckelberg mechanism. In the minimal $B-L$ extension of MSSM, one
 finds that under the assumption of the universality of soft scalar masses, 
charge conservation and in the absence of  a Fayet-Iliopoulos D-term, 
R-parity does not undergo spontaneous breaking by renormalization groups effect~\cite{ffn}.\\

\section{Dark Matter} 
 
{\it Dark matter in SUGRA  unification:}
While in MSSM the LSP  with R parity is stable,  there is no reason that the LSP
would be a neutral particle much less a neutralino.
However, in mSUGRA with universal boundary conditions at the GUT scale, the neutralino turns out 
to be the LSP over a large part of the parameter space under constraints of color and charge
conservation. Thus in mSUGRA the neutralino becomes a candidate for cold dark matter. 
Many extended SUGRA models, i.e., models with non-universalities,  also exhibit the same
phenomenon. 
Stringent constraints have been placed on the allowed lower limit of the neutralino mass
in MSSM/SUGRA models~\cite{Feldman:2010ke}. 
An important test of neutralino dark matter 
will come from  direct detection experiments for the search for dark matter which measure 
 the spin-independent neutralino-proton cross section (see, e.g., ~\cite{Ellis:1987sh}). 
The dual constraints from dark matter searches and from searches for supersymmetry at 
colliders  help delineate stringently the
parameter space of models (see, e.g., ~\cite{Feldman:2011me}).
Recent experiments have made significant progress
in increasing the sensitivity of experiments for this cross section (see, e.g., ~\cite{xenon}). 
Further, some recent experiments
have indicated the possibility of the spin-independent cross section as large as $10^{-40}$cm$^2$
in the low mass neutralino region as low as 5-10 GeV~\cite{Aalseth:2010vx}. 
 However, while 
a  low mass neutralino, as low as 5-10 GeV, is allowed by REWSB, the constraints from WMAP, 
from $b\to s\gamma$ and from $B_s\to \mu^+\mu^-$ essentially eliminate this
 region of the parameter space.
Further, new stringent constraints on dark matter 
arise from the recent data  on the search for 
supersymmetry  ~\cite{cmsREACH,AtlasSUSY,atlas165pb,atlas1fb} 
and for the search for 
 the Higgs boson~\cite{dec13}. In ~\cite{Akula:2011aa} 
 an analysis was carried out including the lower limit  constraints on sparticles masses  and also 
  restricting   the Higgs mass range
to  $123$ GeV to $127$ GeV which appears to be the prime region for the
discovery of the Higgs boson in view of the recent  LHC data ~\cite{dec13}.
In this case after the 
imposition of the 
 radiative electroweak symmetry breaking, relic density
and the FCNC constraints  one finds that essentially all of the  mSUGRA parameter 
points that give a  $123-127$ Higgs boson mass  produce a proton-neutralino spin-independent cross section 
that lies  just beyond the most recent  experimental limits  from 
the XENON collaboration~\cite{xenon}.  However,  it is very encouraging that most of this region would be 
 within   the projected reach of  
 XENON-1T~\cite{futureXENON} and SuperCDMS~\cite{futureSCDMS}.  \\

{\it Cogenesis:}
Another interesting aspect of dark matter concerns cosmic coincidence, i.e., 
that the baryonic matter and the non-baryonic  dark matter
are of the same size, and   more precisely ~\cite{wmap}
$\Omega_{DM}/{ \Omega_B} = 4.99 \pm 0.20$.
The above appears to indicate that these two types of matter
are somehow related and have the same origin. 
Though the topic is outside the main theme of this talk, it is of current interest and we 
give a brief discussion of it here. Thus  there is  the  suggestion 
 of the so called asymmetric  dark 
 matter which proposes that dark matter 
could be created by a transfer of  a  net $B-L$ asymmetry
created in the early universe to some standard  model singlet which would be the dark matter candidate
~\cite{Davoudiasl:2012uw}.
There are two main issues that need attention here.  The first is how one may carry out such a transfer and the second how
the dark matter produced by thermal processes (symmetric dark matter)  is dissipated. 
Regarding the first item,  a transfer  of $B-L$  can occur via interactions of the type~\cite{Kaplan:2009ag}
$\frac{1}{M_a^n} O_{DM} O^{SM}_{asy}$,
where $O^{SM}_{asy}$ is constituted of only the standard model particles and carries a net $B-L$ quantum number 
and $O_{DM}$ is constituted of dark matter fields and carries the opposite $B-L$ quantum number. 
Regarding the second item, one needs to demonstrate in a quantitative fashion that the symmetric 
component of dark 
 matter is efficiently annihilated. 
We do not speculate on how the $B-L$ asymmetry originates and simply assume that it exists. 
Assuming a pre-existing $B-L$ the analysis then revolves around a transfer of the $B-L$ from
the visible to the dark matter sector. A common assumption is that 
the $B-L$ transfer occurs at a temperature $T >T_c$, where $T_c$ is the temperature
of electroweak phase transition, and also  $T> T_{sph}$, where $T_{sph}$ is the sphaleron temperature
(although a variety of other possibilities have also been explored).
There are various possibilities for $O_{asy}^{SM}$ and for $O_{DM}$. Thus, e.g., $O_{asy}^{SM}$ could
be $LH, (LH)^2, LLE^C, LQD^C, U^CD^CD^C$ while $O_{SM}$ could be $\psi^k$ where $k>2$
where $\psi$ is the dark matter particle. 
In the specific  analysis  of ~\cite{Feng:2012jn} interactions involving the leptonic doublets are considered 
and  no flavor mixing is assumed in this sector so $L_i$ are separately conserved. Further, in SM
we can gauge one combination of $L_i-L_j$ (or $B-L$) ~\cite{He:1991qd} which we choose to be $L_{\mu} -L_{\tau}$. 
The gauging is done by the \st mechanism~\cite{kn12,FLNPRL,kctc,Liu:2011di}.
Using  equilibrium conditions for the 
chemical potentials we calculate the ratio $n_{DM}/n_B$ and then one has 
$
\frac{\Omega_{\rm DM}}{ \Omega_{\rm B}} = \frac{m_{\rm DM} n_{\rm DM}}{m_{\rm B} n_{\rm B}} = 4.99\pm 0.20
$
which is satisfied for $m_{DM}\leq 20$ GeV for a variety of models.
The main constraint on the model arises  from $g_{\mu}-2$. 
As mentioned already 
an important element in achieving a successful model of asymmetric dark matter
is to have an efficient mechanism for the annihilation of  its  symmetric component. This is achieved in the
proposed 
 model by resonant annihilation via the Breit-Wigner $Z'$ pole. 
~\cite{Griest:1990kh,Nath:1992ty}.
 The $Z'$ does not couple with the first generation leptons nor with the quarks, and couples only with
 the second and third generation leptons.  Thus the usual LEP constraints on the ratio $M_{Z'}/g$ do 
 not apply to the $Z'$ we consider  and consequently the $Z'$ here can have a mass close to twice the mass
 of the dark matter particle and a rapid annihilation of the symmetric component of dark matter
 can occur. A direct extension of the above mechanism can also be carried out for the supersymmetric
 case~\cite{Feng:2012jn}. Here, however, a new feature arises in that in addition to the asymmetric 
 dark matter one also has another candidate for dark matter, i.e., the neutralino and thus one has 
 a two component dark matter picture. In order for the asymmetric dark matter mechanism to work 
 one must deplete the neutralino component so that it is no more than, say one tenth of the asymmetric
 dark matter component. This was accomplished in ~\cite{Feng:2012jn}. Now the asymmetric dark matter
 is not observable  in direct detection experiments because it has no direct interaction with quarks.
  It is then interesting to ask if  the suppressed neutralino component may be measurable in experiment.
  Indeed the analysis given in   ~\cite{Feng:2012jn} shows that even a subdominant neutralino will
  be accessible in future direct detection experiments such as future XENON-1T~\cite{futureXENON} and
   superCDMS~\cite{futureSCDMS} experiments.

\section{Conclusion}
The $SO(10)$ model continues to be an attractive framework for unification.
However,  in $SO(10)$ model building one is faced with many choices for the Higgs fields.
Thus typically more than one representation is needed to break the GUT symmetry
which makes the model less predictive as  one needs
additional assumptions on the VEVs of the Higgs fields 
to explain gauge coupling unification.  
This issue is resolved in models using higher Higgs representations such as $144+ \overline{144}$ 
or $560 +\overline{560}$. 
Further, the latter model anchored in $560+\overline{560}$ also gives a natural doublet-triplet 
splitting via the missing partner mechanism. 
The GUT group embedded in  supergravity, i.e., supergravity grand unification,  
allows one to make contact between GUT 
physics and low energy physics.  Specifically the SUGRA GUT model predicts the light Higgs boson mass to be
below $\sim$130 GeV.
In this context the recent LHC data on the Higgs boson is very 
encouraging. The experimental data gives a hint of the light Higgs boson mass in the region around 125 GeV. 
Further, the analysis within mSUGRA indicates that a Higgs boson in this mass range requires sparticles to be
generally heavy. It is found that most of the parameter space of models lies on the Hyperbolic Branch ~\cite{Chan:1997bi,HB2} of radiative breaking of the electroweak symmetry and specifically on 
Focal Surfaces~\cite{Akula:2011jx}.
However, several sparticles could still be relatively light including the light stop and
the light sbottom, as well as the charginos and neutralinos while the gluino mass can still lie below 1 TeV. 
 More data expected from the LHC  in the coming months will  provide further tests of SUGRA GUTs.
Thus LHC is an important laboratory for the test of both SUSY and GUTS and more specifically of 
the SUGRA GUT model.\\

\noindent
{\bf Acknowledgments}\\
This research is supported in part by the U.S. National Science Foundation (NSF) grants
PHY-0757959 and PHY-0969739  and through XSEDE under grant number TG-PHY110015.

\end{document}